\documentclass[twocolumn,floats,aps,amsmath,showpacs,amssymb]{revtex4}

\usepackage{graphicx}
\usepackage{epsfig}
\newcommand{\eqa}{\begin{eqnarray}}
\newcommand{\ena}{\end{eqnarray}}

\newcommand{\bra}{\langle}
\newcommand{\ket}{\rangle}

\newcommand{\eq}{\begin{equation}}
\newcommand{\en}{\end{equation}}

\begin{document}

\vskip 0.4cm

\title{$k-$strings and baryon vertices in $SU(N)$ gauge theories}

\author{Ferdinando Gliozzi}

\affiliation{ Dipartimento di Fisica Teorica
        dell'Universit\`a di Torino, and\\ 
       INFN, Sezione di Torino, via P.Giuria 1, I-10125 Torino, Italy }    
\begin{abstract}
It is pointed out that the sine law for the $k-$string tension emerges as the 
critical threshold below which the spatial $Z_N$ symmetry of the static baryon 
potential is spontaneously broken. This result applies not only to  
$SU(N)$ gauge theories, but to any gauge system with stable $k-$strings 
admitting a baryon vertex made with $N$ sources in the fundamental 
representation. Some simple examples are worked out.  
\end{abstract}
\pacs{11.15.-q, 12.38.Aw, 11.15.Ha, 12.39.Pn}
\maketitle
The linear rising of the interquark potential in $SU(N)$ gauge theories
suggests that  the chromo-electric flux tube between static quarks in
 the fundamental representation is
 localised in a thin tube joining those charges and it is widely believed 
that the long-distance physics of such flux tubes is given by an effective 
string theory.

In addition to charges in the fundamental representation one can consider 
the potential between static charges in higher representations of the gauge 
group. This has been the subject of
 considerable theoretical attention since mid-1970's, when it was 
pointed out \cite{ks} that the strong coupling limit or dimensional 
reduction arguments \cite{aop} suggest the 
Casimir scaling law, {\sl i.e.} the hypothesis that the string 
tension for a given representation is proportional to the quadratic 
Casimir operator.   
This seems to describe accurately the potential between sources in different
 representations extracted from numerical studies of $SU(3)$ lattice gauge 
theory at intermediate distances \cite{{sd},{ba},{ss}}. 

Note however that the long-distance properties of the flux tube attached  
to a  charge built up of $j$ copies of the fundamental representation 
should depend only on its $N$-ality  $k\equiv j~({\rm mod}\,N)$,
the reason being that all representations with the same $k$ 
can be converted into each other by the emission of a suitable number of 
soft gluons.   As a consequence, the heavier strings of given $N$-ality $k$
are expected to decay into the string with smallest string tension
\cite{as,me}.  
The corresponding string is usually referred to as a $k$-string. If its 
tension $\sigma_k$ satisfies the inequality $\sigma_k<k\,\sigma$, where
$\sigma\equiv\sigma_1$ is the  tension of the fundamental string, 
 the  $k-$string is stable against 
decay into $k$ 1-strings.
Charge conjugation implies $\sigma_k=\sigma_{N-k}$, therefore $SU(N)$ has only
 $[N/2]$ distinct $k$-strings. 

Stable $k$ string are expected to belong 
to the antisymmetric representation with $k$ quarks. This is also supported 
by Casimir scaling, which in this case yields 
\eq
\sigma_k^{(c)}=\sigma\frac{k(N-k)}{N-1}~.
\label{casi}
\en
In the large $N$ limit it has been argued \cite{as} that 
$\sigma_k=k\,\sigma+O(\frac1{N^2})$
which seems to exclude Casimir scaling law as an exact formula.

Another competing hypothesis is the sine law:
\eq
\sigma_k^{(s)}=\sigma\frac{\sin(k\pi/N)}{\sin(\pi/N)}~,
\label{sine}
\en
which  has been found in ${\cal N}=2$ supersymmetric $SU(N)$ gauge
theory softly broken to ${\cal N}=1$ \cite {ds}, in the M theory 
description of  ${\cal N}=1$ supersymmetric $SU(N)$ gauge theory, called MQCD
\cite{hsz} and, more recently, in the AdS/CFT framework \cite{hk}. In some 
cases this formula is expected to be exact, while in others 
 perturbative corrections have been found \cite{ak}.
 
Lattice calculations in pure $SU(N)$ gauge models for $N=6$
\cite{dprv} and $N=4,5,6,8$ \cite{lt,ltw} in $D=3+1$ point to the $k-$string 
tensions lying partway between the Casimir scaling and the sine law, however
there is no complete consensus and some dedicated studies favour the sine 
formula \cite{dprv2}.
Enlarging the analysis to other gauge groups, we shall see two instances of 
 $Z_4$ gauge models  where the 2-string tension can be 
exactly evaluated in any dimension  in an almost obvious way. In one case,
as it turns out, $\sigma_2=2\sigma$ while in the other    
$\sigma_2=\sigma$ (or, more generally, $\sigma_k=\sigma$ for a $Z_N$ 
gauge group) and there are reasons to believe  that 
there exists a set of gauge models which continuously interpolate 
between these two extremal values. Nonetheless, Eq.(\ref{sine}) plays a 
special role: in this Letter we show that in whatever gauge theory in 
which the  center of the gauge group is $Z_N $ the $k-$string tension given by 
the  sine law has a simple geometrical  meaning: it is the threshold 
below which  the spatial $Z_N$  symmetry of the baryon vertex is 
spontaneously broken.

A baryonic vertex is a gauge-invariant coupling of $N$ multiplets in the
fundamental representation which gives rise to finite energy configurations
with $N$ external quarks \cite{xa,sw}. When the separations among these 
quarks is large,
one expects that $N$ strings of chromo-electric flux form, which meet at 
a common junction; their world-sheet forms a $N-$bladed surface with a common 
intersection. Due to its shape in the case of SU(3), this description of the 
baryonic potential is known as the Y-Ansatz. A different description, known 
as $\Delta-$Ansatz, follows from the assumption \cite{co} that  the two-body 
interaction is the relevant one for any $SU(N)$. Its name comes from the 
linear rise of the potential with the perimeter of the triangle in the $SU(3)$
case. Lattice data for  
$SU(3)$ \cite{{adt},{adj},{tt}} seem to support the 
$\Delta$- Ansatz at 
short distances. At interquark separations larger than $\sim\,0.8$ fm the 
$\Delta$- Ansatz breaks down and there is a gradual transition to the Y-Ansatz.

In the mapping from four-dimensional gauge theories to string theory in 
{\sl AdS} space \cite{ma} a $SU(N)$ baryon vertex has been explicitly obtained 
 by wrapping a fivebrane over $S^5$ \cite{wi}.

\begin{figure}[t]
\begin{center}
\mbox{~\epsfig{file=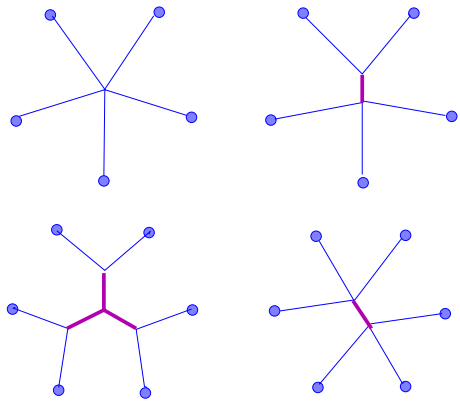,width=6.0cm}}
\vskip .2 cm

\caption{Baryon vertices. The two top diagrams depict the symmetric 
 and the broken symmetry configurations of the $SU(5)$ baryon vertex,
the two bottom diagrams show  two symmetry breaking 
schemes in the $SU(6)$ case.  The solid dots represent the 
quark sources in the fundamental representation, The thin lines 
denote the fundamental strings, while the  thick lines
are 2- and 3-strings. }
\label{Figure:1}  
\end{center}
\end{figure}
 When $N>3$ one has to envisage the possibility that $k$ neighbouring strings
of the baryon vertex coalesce into a single $k-$string
\cite{adt}. Note that the 
external quarks belong to a fully antisymmetric combination, which is the most 
favourable condition to $k-$string formation. What is the cost in energy of 
such a configuration?  To answer this question, we place external quarks 
at the $N$ corners  of a regular polygon  inscribed in a circle 
of radius $R$. When $R$  is much larger than the string formation scale  
the gauge flux is squeezed into $1$-strings attached to the quarks. 
Preparing the external charges in a spatially $Z_N$-symmetric 
configuration does not necessarily imply that the structure of the gauge 
flux preserves $Z_N$: allowing strings to coalesce  
requires breaking such a symmetry. Assuming $k$-strings do not form leads 
to a gauge flux  concentrated in a $N$-bladed string 
world-sheet bounded by the parallel world-lines of the static quarks. 
In this case  the configuration of minimal energy is  
$Z_N$-symmetric. The baryon potential $V_N$ is roughly  proportional to 
the total length of the strings, therefore one has $V_N=N\,\sigma\,R+
O(\frac1R)$, 
where the leading $O(\frac1R)$ correction is due to the Casimir 
energy \cite{{lu},{jf}}. 

\begin{figure}[tb]
\begin{center}
\mbox{~\epsfig{file=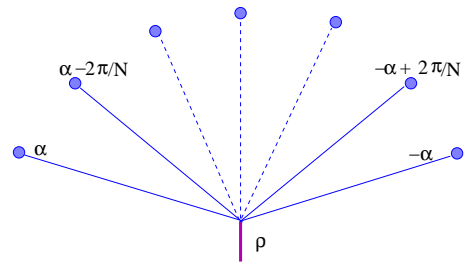,width=6.0cm}}
\vskip .2 cm
\caption{Formation of a $k-$string inside a baryon vertex. }
\label{Figure:2}  
\end{center}
\end{figure}

The formation of $k-$strings breaks the 
spatial $Z_N$ symmetry. Some examples of these configurations are drawn in 
Fig.\ref{Figure:1}. 
More general symmetry-breaking schemes can be encoded in an arbitrary 
partition of $N=k_1+k_2+\dots+k_m$,
 where  $k_a$ is the number of neighbouring  
strings which coalesce into a single $k_a$-string. The associated string 
configuration is generated by iterating  the basic motif depicted in 
Fig.\ref{Figure:2}. Denoting by  $Re^{i\frac{2\pi j}{N}}$ $(j=1,2\dots N)$
the position of the vertices in the polygon, we can generalise 
the static potential to
\eq
V_{\{k_1,k_2,\dots k_m\}}=\sigma \,R\,\sum_{a=1}^mf_{k_a}(\rho_a)+
O\left(\frac1R\right)~,
\label{pot}
\en
where $\rho_a\, R$ is the length of the $k_a$-string and
\eq
f_k(\rho)=\rho\,\frac{\sigma_k}\sigma+\sum_{j=0}^{k-1}\left\vert\rho-
e^{i\frac{2\pi j}N-i\alpha}\right\vert~,
\label{motiv}
\en
with $\alpha=\frac{k-1}N\pi$ (see Fig.\ref{Figure:2}). This choice of 
$\alpha$   eliminates any dependence on the orientation of the  
$k-$string. The Ansatz (\ref{pot}) is based on the assumption
that the common junction of the strings (which is the symmetry axis of the 
configuration) is not displaced by  $k-$string formation. This is 
obviously true as long as the system preserves  a residual   symmetry,
but it is also justified for more general string breaking schemes, owing to 
the fact that we are interested in the threshold of the symmetry breaking.

In order not to break the $Z_N$ symmetry the configuration 
of minimal energy should be characterised by $\rho_{k_a}=0$ for 
all $a=1,\dots,m$.

First, from the observation that $g(\rho)=\rho+\vert\rho-e^{i\theta}
\vert$ is always increasing and $g''(\rho)>0$ for any $\rho$ and $\theta$, 
one shows at once that the function $f_k(\rho)$ has only one minimum 
in the whole $\rho$ range. Then, 
Taylor expanding $f_k(\rho)$ around $\rho=0$ yields
\eq
f_k(\rho)=k+\rho\,\frac{\sigma_k-\sigma_k^{(s)}}{\sigma}+\frac{\rho^2}4
\left(k-\frac{\sin \frac{2\pi k}N}
{\sin\frac{2\pi} N}\right)+\cdot\cdot\cdot
\label{taylor}
\en
where the crucial sine ratio $\frac{\sigma^{(s)}_k}{\sigma}$ in the linear 
term arises from the geometric sum
\eq
\sum_{l=0}^{k-1} e^{i\frac{2\pi l}{N}-i\alpha}=\frac{e^{-i\alpha}- 
e^{-i\alpha+i\frac{2\pi k}{N}}}{1- e^{i\frac{2\pi}{N}}}  
=\frac{\sin\frac{\pi k}N}{\sin\frac{\pi}N}\equiv\frac{\sigma^{(s)}_k}{\sigma}~.
\en

Assuming $\sigma_k=\sigma_k^{(s)}$ for all allowed values of $k$
yields the three relationships
\eq
f_k'(0)=0~,~f_k''(0)>0~,~V_{\{k_1,k_2,\dots k_m\}}=V_N~,
\en
which tell us that  for $\sigma_k\ge\sigma_k^{(s)}$ 
the  $Z_N$-symmetric baryon vertex should be stable against the formation 
of $k$-strings, while for $\sigma_k<\sigma_k^{(s)}$ this is no longer true 
and the system breaks up into less symmetric configurations made with 
$k$-strings. In other terms, the sine law is the critical threshold below
which the $Z_N$ symmetry of the baryon vertex is spontaneously broken.  

In $3+1$ dimensions one can  place the static quarks  in 
non-planar configurations, but only in the special cases $N=4,6,8,12,20$
one can arrange them in a fully symmetric configuration, 
corresponding to the vertices of the platonic solids. The thresholds of 
symmetry breaking are in these cases lower than the corresponding $Z_N$ 
values. For instance, for the tetrahedron 
we get $\frac{\sigma_2}{\sigma}=\frac2{\sqrt{3}}$.   
\begin{figure}[tb]
\begin{center}
\vskip -.8 cm
\mbox{~\epsfig{file=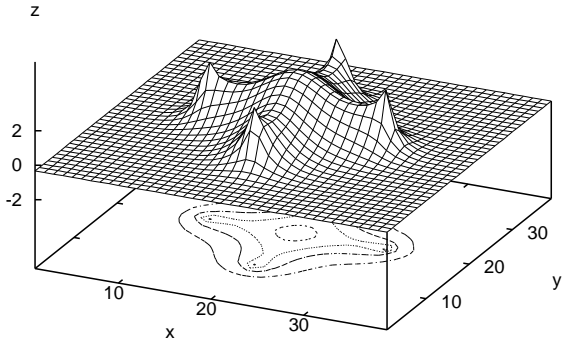,width=9.8cm}}
\caption{The shape of the flux tube in a slice of a $Z_4$ static baryon 
described by the model (\ref{zz}). The measured quantity is the plaquette. 
The static charges are placed at the 
corners of a square of side $L/a=10\sqrt{2}$ in a 
$16\times64\times 64$ lattice at $\beta=1.446\,$ corresponding to 
$\sigma a^2\simeq 0.05$. As expected, no sign of symmetry breaking is 
observed. }
\label{Figure:3}  
\end{center}
\end{figure}
\begin{figure}[tb]
\begin{center}
\vskip -.8cm
\mbox{~\epsfig{file=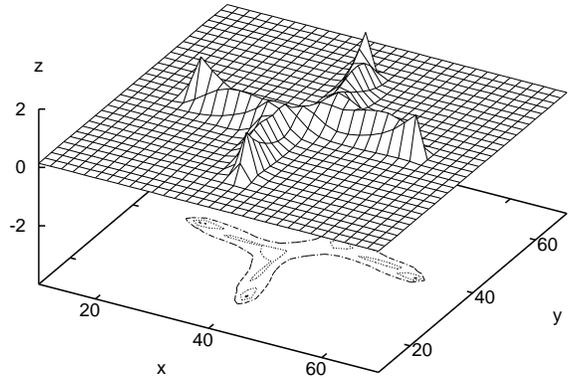,width=9.8cm}}
\caption{The shape of the flux tube in a slice of a $Z_4$ static baryon 
of the model (\ref{zp}). The measured quantity is the plaquette. 
The static charges are placed at the 
corners of a square of side $L/a=20\sqrt{2}$ in a 
$16\times128\times 128$ lattice at $\beta=1.51$ corresponding to
 $\sigma a^2\simeq 0.76$. It is evident the 
formation of a 2-sting  breaking the $Z_4$ symmetry. }
\label{Figure:4}  
\end{center}
\end{figure}

Notice that if $N/k$ is an  integer, there is  a new baryon 
vertex coupling $N/k$ external charges lying in the fully antisymmetric 
representation made with $k$ quarks. It follows 
at once that the subset $\sigma^{(s)}_{jk}$ , $(j=2,3,\dots [\frac N{2k}])$ 
constitutes the set of critical thresholds for the spontaneous breaking of 
the $Z_{\frac Nk}$ symmetry of this kind of baryon.

In contradistinction to what happens in MQCD or in other supersymmetric gauge 
theories, lattice calculations in pure $SU(N)$ with $N=4,5,6,8$
\cite{{dprv},{lt},{ltw}} put all these gauge models in the broken symmetry 
phase. The scheme 
of spontaneous symmetry breaking depends on the spectrum of the $k-$string 
tensions. For instance, in 3D $SU(6)$ it turns out that the pattern
$Z_6\to Z_3$ is preferred to $Z_6\to Z_2$.

Numerical methods for revealing the distribution of gauge fields 
within the static baryonic potential are now available \cite{{ib},{ow},{bc}}. 
The question is whether such a spontaneous symmetry breaking effect is 
accessible in realistic simulations: if the size of the system is not large 
enough, we expect that the $SU(N)$ baryonic vertex  undergoes back-and-forth 
tunnelling among the different vacua, obscuring this effect. 
 
Much larger volumes can be reached studying gauge systems with discrete gauge 
groups. Indeed it has to be emphasised that the above considerations 
apply to whatever confining gauge theory admitting a baryonic vertex.
Let us restrict attention to two particularly simple gauge models. 
The partition function of the first model is
\eq
{\cal Z}_{Z_N}(\beta)=\sum_{\phi_\ell\in\{\frac{2\pi j}N\}}
\prod_{P}e^{\beta\cos\phi_P},~\phi_P=\sum_{\ell\in P}\phi_\ell~,
\label{zz}
\en
where $\ell$ and $P$ denote the links and the plaquettes of a $D-$ dimensional
 hypercubic lattice. As in any gauge theory, the most basic observables 
are the Wilson loops $W_k(\gamma)=e^{ik\phi_\gamma}$, where $\gamma$ is a 
closed path of the lattice, $\phi_\gamma=\sum_{\ell\in\gamma}\phi_\ell$, 
and $k$ is the number of units of the fundamental charge of $Z_N$. 

It has been shown long ago \cite{gln} that in the case $N=4$ the above 
gauge theory  is fully equivalent to a $Z_2\times Z_2$ theory in any space 
dimension, namely,
\eq
{\cal Z}_{Z_4}(\beta)={\cal Z}_{Z_2\times Z_2}(\beta/2)=
{\cal Z}^2_{Z_2}(\beta/2)~.
\en 
All the $Z_4$ quantities can be expressed through the $Z_2$ quantities.
In particular
\eq
\bra W_2(\gamma)\ket_{Z_4,\beta}=\bra W_1(\gamma)
\ket^2_{Z_2,\frac\beta2}=\bra W_1(\gamma)\ket^2_{Z_4,\beta}~.
\label{w4}
\en
The confining phase shows up in an area law decay of the vacuum expectation 
value of large Wilson loops, therefore comparing left-hand side 
and right-hand side of Eq.(\ref{w4}) yields $\sigma_2=2\sigma$. Thus, there 
is no stable 2-string and the baryonic vertex keeps its spatial symmetry
in any dimension.

The other class of illustrative models is defined through a simple
modification of  Eq.(\ref{zz}):
\eq
{\cal Z}'_{Z_N}(\beta)=\sum_{\phi_\ell\in\{\frac{2\pi j}N\}}
\prod_{P}e^{\beta\,\delta_{\phi_P,0}}~~,
\label{zp}
\en
where now the  plaquette variable is given by the  Kronecker delta 
$\delta_{\phi,0}$, which  is 1 only if $\phi\equiv 0$ modulo $2\pi$. Sewing 
together a proper number of plaquettes one can generate any Wilson loop, 
which inherits the same property, {\sl i.e.}
$\,W_k(\gamma)=\delta_{k\phi_\gamma,0}$. Since all the configurations with 
$W_1(\gamma)=1$ imply $W_{k>1}(\gamma)=1$, it follows that
$\bra W_k(\gamma)\ket\ge \bra W_1(\gamma)\ket$ with $k>1$.
Arguing as above, taking into account that $\sigma_k\ge\sigma$, now we 
get $\sigma_k=\sigma$, so that 
$\sigma_k<\sigma^{(s)}_k$  for any $N$ and in any dimension. As a consequence,
 the spatial $Z_N$ symmetry of the baryonic vertex is 
spontaneously broken. As $N$ increases, the pattern of symmetry breaking 
may become rather involved.
As an example, using Eq.s (\ref{pot}) and (\ref{motiv}) one can see that 
a $Z_{12}$ symmetric baryon vertex breaks first in four 3-strings which 
in turn break down into two 6-strings. 

In order to see the distribution of the gauge field inside a baryon vertex 
we simulated the  $Z_4$ gauge models defined in Eq.s (\ref{zz}) and 
(\ref{zp}) in $D=2+1$. We used the plaquette as a probe in the vacuum 
modified by the presence of four static sources  placed at the corners 
of a large square of side $L$. They 
are  represented by four parallel Wilson lines wrapped around a 
periodic direction. To reach the required large distances, the Monte Carlo 
simulations were  actually performed in the dual versions of these 
systems, which are  simple spin models: the 
system (\ref{zz}) can be exactly mapped  into  the clock $Z_4$ model and the 
system  (\ref{zp}) in the 4-state Potts model. This choice allows one to use 
 efficient nonlocal cluster simulation algorithms \cite{Sw}. The resulting 
flux-tube profile is shown in Fig.\ref{Figure:3} and in Fig.\ref{Figure:4}.
In the former the spatial $Z_4$ symmetry of the baryon vertex is preserved, 
while in the latter the formation of a 2-string is clearly visible.
Its length $\ell$ is about 40\% of the side $L$, according to  
the configuration which minimises Eq.(\ref{pot}), yielding, in this case, 
$L-\ell=L/\sqrt{3}$.    

In conclusion, in this work it has been found a simple physical interpretation 
of a behaviour of $k-$strings -the sine law- which so far was regarded as 
a mathematical consequence of supersymmetry. Here it has been instead related 
to the marginal stability of  $Z_N$-symmetric baryon vertices.
It would be very interesting to discover some relation between these two 
approaches.

\end{document}